\documentstyle[12pt]{article}
\input{psfig.tex}
\newcommand{\etal}{{\em et\ al.}\ }

\newcommand{\kms}{km s$^{-1}$\ }

\newcommand{\chisq}{${\chi}^{2}$}
\newcommand{\chinusq}{${\chi}_{\nu}^{2}$}

\def\lesssim{\mathrel{\hbox{\rlap{\hbox{\lower4pt\hbox{$\sim$}}}\hbox{$<$}}}}
\def\gtrsim{\mathrel{\hbox{\rlap{\hbox{\lower4pt\hbox{$\sim$}}}\hbox{$>$}}}}

\begin{document}

\vspace{0.2in}

\centerline{                       THE ABSOLUTE LUMINOSITIES OF}
\centerline{                   THE CAL\'{A}N/TOLOLO TYPE Ia SUPERNOVAE}

\vspace{0.3in}
\centerline{                               Mario Hamuy$^{1,2}$}
\centerline{                              M. M. Phillips$^1$}
\centerline{                            Robert A. Schommer$^1$}
\centerline{                           Nicholas B. Suntzeff$^1$}
\centerline{                                 Jos\'{e} Maza$^{3,4}$}
\centerline{                                 R. Avil\'{e}s$^1$}
\vspace{0.3in}
\noindent $^1$ National Optical Astronomy Observatories$^*$,
Cerro Tololo Inter-American Observatory, Casilla 603, La Serena, Chile\\

\noindent $^2$ University of Arizona, Steward Observatory, Tucson,
Arizona 85721\\

\noindent $^3$ Departamento de Astronom\'{i}a, Universidad de Chile, Casilla 36-D,
Santiago, Chile\\

\noindent $^4$ C\'{a}tedra Presidencial de Ciencias (Chile), 1996-1997.\\

\noindent electronic mail: mhamuy@as.arizona.edu, mphillips@noao.edu, rschommer@noao.edu,\\
\noindent nsuntzeff@noao.edu, jmaza@das.uchile.cl\\
\vspace{0.1in}

\noindent Running Page Head : ABSOLUTE LUMINOSITIES OF TYPE Ia SNe\\

\noindent Address for proofs:  M. M. Phillips, CTIO, Casilla 603, La Serena, Chile\\

\vspace{0.1in}

\footnoterule

\vspace{0.1in}

\noindent $^*$Cerro Tololo Inter-American Observatory, National
Optical Astronomy Observatories, operated by the Association of Universities
for Research in Astronomy, Inc., (AURA), under cooperative agreement with
the National Science Foundation.\\

\eject
\centerline{                        ABSTRACT}

We examine the absolute luminosities of 29 SNe Ia in the Cal\'{a}n/Tololo
survey. We confirm a relation between the peak luminosity of the SNe and
the decline rate as measured by the light curve, as suggested by Phillips
(1993). We  derive linear slopes to this magnitude-decline rate relation in
BV(I)$_{KC}$ colors, using a sample with B$_{MAX}$-V$_{MAX}$ $<$ 0.$^m$2. The
scatter around this linear relation (and thus the ability to measure SNe Ia
distances) ranges from 0.$^m$13 (in the I band) to 0.$^m$17 (in the B band).
We also find evidence for significant correlations between the absolute magnitudes or
the decline rate of the light curve, and the morphological type of the
host galaxy.

\eject

\section{  Introduction}
     Given the high intrinsic luminosities of type Ia supernovae (SNe Ia hereafter),
considerable effort has been devoted during recent years to evaluating the usefulness of these
objects as extragalactic distance indicators.  Modern, high-quality photometry obtained with
CCD detectors has revealed that SNe Ia display a wide range ($>$1 mag) of intrinsic optical
luminosities at maximum light (e.g., Phillips 1993; Hamuy \etal 1995, hereafter referred to as
Paper IV), potentially complicating the use of these  objects as standard candles.  There is
also mounting evidence that the light curves of SNe Ia vary significantly in their shapes, and
that these variations are correlated with the intrinsic brightness at maximum light (Phillips \etal
1987; Maza \etal 1994, hereafter Paper II; Paper IV; Suntzeff 1996).

     More than 20 years ago, Barbon, Ciatti \& Rossino (1973) suggested that SNe Ia could
be divided into ``fast'' and ``slow'' decliners, depending on the initial decline rate of the B~light
curve.  Pskovskii (1977, 1984) took this idea a step further, suggesting that the initial decline rates
of SNe Ia were correlated with their intrinsic luminosities, with slow decliners being
intrinsically brighter than fast decliners.  Unfortunately, these results remained uncertain due
to the relatively poor quality of the (mostly photographic) photometry available at that time
and the related problem of contamination of these measurements by the underlying light of
the host galaxies (see Boisseau \& Wheeler 1991).  In addition, the samples of Barbon \etal
and Pskovskii both included SNe which we now recognize to be type Ib/Ic events.  Thus, the
existence of a peak luminosity-decline rate relationship was widely doubted until three years
ago, when Phillips (1993) reexamined the  issue using a subsample of nine
nearby, well-observed 
SNe Ia.  With surface brightness fluctuations (SBF) and Tully-Fisher (T-F) distances
available for these SNe, Phillips showed that the initial decline rate of the B light curve was,
in fact, tightly correlated with the intrinsic B, V, and I maximum luminosity.  This finding
has significant implications for the use of this class of objects as extragalactic distance
indicators since the initial decline rate could, in principle, be used to correct the apparent peak
luminosity (in analogy to the period-luminosity relationship for Cepheid variables).

     The Cal\'{a}n/Tololo SN survey, which was initiated at CTIO with the aim of discovering
distant SNe (Hamuy \etal 1993a hereafter referred to as Paper I), yielded 32 new SNe Ia in
the redshift range 0.01 $\lesssim$ z $\lesssim$ 0.1 during the period 1990-93.  
Through the collaboration of
many visiting astronomers, high-quality CCD light curves were obtained for nearly all of
these discoveries.  These SNe provide an independent opportunity to test the reality of the
decline rate/intrinsic brightness relationship since they are at sufficiently large redshifts that
the radial velocities of their host galaxies serve as an accurate indicator of their relative
distances -- and, hence, their relative luminosities.  A preliminary study in the B and V bands
was reported in Paper IV for 13 Cal\'{a}n/Tololo SNe Ia which confirmed in general terms the
results of Phillips, although the slope of the relationship was less steep.  With
the entire Cal\'{a}n/Tololo database now reduced, a new analysis of this relationship seems in
order.  In this paper we report our definitive results based on light curves in the B, V and I
bands for the 27 best-observed Cal\'{a}n/Tololo events plus 2 additional distant SNe Ia for which
similar data were obtained in our program.  After a brief description of our sample (Sec. 2), we present in
Sec. 3 a final version of the absolute magnitude vs. initial decline rate diagram for the
Cal\'{a}n/Tololo events.  This is compared in the same section with the luminosity-decline rate
relation observed for an updated sample of nearby SNe Ia with SBF, 
Planetary Nebula Luminosity Function (PNLF), and Cepheid
distances.  Finally, in Sec. 4, both of these samples are used to reexamine the possibility
(suggested in Paper IV) that the most luminous SNe Ia occur preferentially in galaxies having
younger stellar populations.  In accompanying papers, we present and discuss the Hubble
diagram for the full set of Cal\'{a}n/Tololo SNe Ia (Hamuy \etal 1996a, hereafter Paper VI)
and publish the individual BV(I)$_{KC}$ light curves (Hamuy \etal 1996b, hereafter Paper VII).

\section{  The Samples}

\subsection{                           Distant Sample}

     Of the 32 SNe Ia discovered during the course of the Cal\'{a}n/Tololo
survey, adequately sampled
BV(I)$_{KC}$ light curves were obtained for a total of 27 events.  To this sample we have added two
SNe Ia -- 1990O and 1992al -- which, although not found by us, were included in our
program of followup photometry.  Hence, the total sample of distant SNe Ia considered in this
paper is 29.

     Maximum-light magnitudes were measured directly from the observations whenever
possible. We were able to perform this measurement for 11 SNe for which the time of the first photometric
observation was no later than day +1 (counted since the peak of the B light curve), by
fitting a low-order (3-4) polynomial to the data around maximum light.
Among these 11 SNe, five of them possess well sampled light curves starting even a
few days before peak and spanning through day +15, allowing thus a direct measurement
of the initial decline rate of the B light curve. We performed this measurement by
fitting higher order (5-6) polynomials to the data gathered
during the rising branch and the initial decline phase.
For the majority of the events whose light curves were not sufficiently well-sampled, or for which the
follow-up observations did not begin until after maximum light, the peak magnitudes and
initial decline rates were estimated using a \chisq-minimizing fitting procedure similar to that
described in Paper IV.  Note, however, that in order to include the I-band data, an updated
and expanded set of BV(I)$_{KC}$ light curve templates were employed.  This new set of templates
are the subject of Paper VIII in this series (Hamuy \etal 1996c).
Further details of this fitting procedure are given in paper VII.

     Table 1 lists the 29 SNe and the relevant information for this study in the following
format:\\
\noindent Column (1): SN name\\
\noindent Column (2): the ``color'' of the SN, B$_{MAX}-$V$_{MAX}$.\\
\noindent Column (3), (4) and (5): the absolute B, V, and I peak magnitudes of
the SN calculated from the apparent magnitude, the redshift of the host galaxy
(in the CMB frame), and an assumed
Hubble constant of H$_\circ$= 65 \kms Mpc$^{-1}$. As we show in Paper VI,
this value corresponds to the Hubble diagram of the distant SNe Ia with its
zero point duly calibrated with Cepheid distances to nearby SNe Ia.
These magnitudes were corrected for foreground
extinction in the direction of the host galaxy (Burstein \& Heiles 1982), and for the K terms
calculated by Hamuy \etal (1993b). { \it Note, however, that no correction has been applied for
possible obscuration in the host galaxy (see Sec. 3)}.\\
\noindent Column (6): the decline rate parameter $\Delta$m$_{15}$(B), defined by
Phillips (1993) as the amount in magnitudes that the B light curve decays in
the first 15 days after maximum light. In most cases (24 SNe) this parameter was estimated
through the template fitting procedure previously described in Hamuy \etal 1994 (hereafter
referred to as Paper III) and paper IV.\\
\noindent Column (7): the B--V color of the host galaxy.\\
\noindent Column (8): the morphological type of the host galaxy following the
classification scheme in The Hubble Atlas of Galaxies (Sandage 1961).

\subsection{                             Nearby Sample}

     The sample of nearby SNe Ia employed in this paper is similar as that
used by Phillips (1993), and consists of those events with 1) precise CCD or photoelectric
photometry, 2) well-sampled light curves with coverage beginning either before or at
maximum light, and 3) accurate relative distances measured for the host galaxy.
The major difference with Phillips (1993) is that we have used only distances
measured via Cepheids or the SBF and PNLF methods since there is growing evidence that
the latter two techniques are on the same zero point as the Cepheids (Tonry 1996;
Jacoby 1996).

     Data for this revised sample are listed in Table 2.  Note that, in several cases, the
absolute magnitudes and decline rates in this table have changed somewhat with respect to
those given by Phillips (1993).  This is due either to the new distance moduli used
(see above), the availability of improved photometry
(SNe 1990N and 1992A), and/or slightly different assumptions of the dust extinction due
to the host galaxy (SN 1986G).  The latter issue is discussed in more
detail in Sec. 3.  The format of Table 2 is as follows:\\
\noindent Column (1): SN name.\\
\noindent Column (2): the host galaxy.\\
\noindent Column (3): the host galaxy morphology following the
classification scheme in The Hubble Atlas of Galaxies (Sandage 1961).\\
\noindent Column (4): the distance modulus of the host galaxy.\\
\noindent Column (5): references to the distance modulus given in column (4).\\
\noindent Column (6): the method used to determine the distance modulus (SBF = surface brightness
fluctuations, PNLF = Planetary Nebulae luminosity function).\\
\noindent Column (7): the decline rate parameter $\Delta$m$_{15}$(B) measured
directly from the B light curve observations.\\
\noindent Column (8): the color excess E(B-V).  Except for SN 1986G, dust
extinction due to the host galaxy has been ignored, and the quoted color excess are the
Burstein \& Heiles (1984) values for interstellar dust in our own galaxy.
With this approach we wish to make the nearby sample reflect
the properties of the distant sample for which we have not attempted to correct
for possible obscuration in the host galaxies.
For the highly reddened SN 1986G, we have estimated the reddening from the
B-V color evolution at late epochs (Lira 1995). Unfortunately, we are not
able to apply this technique for the remaining SNe since the precision
[$\pm0.1^m$ in E(B-V)] is not accurate
enough to determine small reddenings.\\
\noindent Column (9), (10) and (11): the absolute B, V, and I peak magnitudes
of the SN calculated from the apparent magnitude, the distance modulus of the 
host galaxy, and the assumed color
excess.  A standard Galactic reddening law was assumed to convert the color excess to the
extinction in the B, V, and I bands. In the case of SN~1986G, this may not be a good
assumption (see Hough \etal 1987).\\
\noindent Column (12): references to the SN photometry. For SN 1994D we have prefered to use
Smith \etal (1996) data instead of Richmond \etal (1995) because
the former are on our own instrumental system. In any case, the differences
between the two datasets are not greater than 0.05$^m$.\\

\section{  Peak Luminosity-Decline Rate Relation}

     Figure 1 shows the absolute magnitudes of the full sample of 29 distant SNe Ia 
plotted as a function of the decline rate parameter $\Delta$m$_{15}$(B).  The most obvious feature of this
plot is the significant range in intrinsic luminosities displayed by the Cal\'{a}n/Tololo SNe,
amounting to $\sim$2$^m$ in B, $\sim$1$^m$ in V, and $\sim$0.$^m$8  in I.  In general terms, these plots
confirm the results of Phillips (1993) in the sense that SNe with slower decline rates tend to
be intrinsically brighter than fast decliners.  The scatter is clearly greatest in B, and decreases
significantly in the  V and I bands, as might be expected since we have not attempted to correct
for possible dust extinction in the host galaxies.  Ignoring SN 1992K, which is an intrinsically
red object similar to the subluminous SN Ia 1991bg (Paper III),
Table 1 shows that there are only  two SNe in the sample with  B$_{MAX}-$V$_{MAX}$
$>$ 0.20 : 1990Y (0.33$\pm$0.10) and 1993H (0.23$\pm$0.05).  In the case of SN 1993H, narrow Na I
D absorption lines due to interstellar gas in the host galaxy were observed in the spectrum
(equivalent width $\sim$1.2 \AA  ), suggesting that the red color at maximum
light is, indeed, at least
partially the product of uncorrected dust extinction.  The case of SN 1990Y is less clear since
the spectrum obtained by Filippenko \& Shields (1990), which is the only one available to us,
is of such low signal-to-noise that we can only  place an uninteresting upper
limit of 4 \AA \ on
the equivalent width of any possible interstellar Na I D absorption due to the host galaxy. 
For the remainder of this paper, we shall assume that the red color of this SN is also due to
uncorrected dust extinction (since this is the most likely explanation), but this cannot be
proven beyond all doubt with the data at hand.

 With the
three  red SNe excluded (1990Y, 1992K, and 1993H), the resulting range of
colors of the Cal\'{a}n/Tololo SNe becomes significantly narrower (from -0.09 to +0.13). 
This leaves little room for large amounts of extinction in the parent galaxies and
we also believe that, to some extent, some of the scatter in color is intrinsic (eg. SNe 1992bc
and 1992bo; Paper II). 
Note that a color cutoff (at B--V $>$ 0.25)  as an objective criteria to 
exclude contaminated objects was already proposed
by Vaughan \etal (1995).

For the sake of simplicity, we have fitted straight lines to the data of
Fig.1. We have carried out weighted, linear, least-squares fits (Press \etal 1992) to the 26 remaining
low-extinction Cal\'{a}n/Tololo SNe, shown
as solid lines in Fig. 1. The parameters resulting from the fits to this
subsample are summarized in  Table 3. In addition we have tested
a cut based on the completeness of the light-curve coverage at early times.
Since  $\Delta$m$_{15}$(B) is a measurement of the decline rate over a short
time interval after peak
luminosity, we choose to exclude all SNe with no photometric measures within
the first 5 days after the peak, thus eliminating 8 additional SNe 
(90T, 91S, 91U, 91ag, 92J, 92au, 92bk, and 93ah).
The parameters derived in this restricted set of SNe (the second solutions in
Table 3)  are insignificantly
different (less than 1$\sigma$ change) from those for the larger sample, and
the reduced chi-squared (\chinusq) values and dispersion around the relation are not improved, so
a restriction to this subsample does not appear to be necessary.
Thus the preferred slopes are 0.784$\pm$0.182 in B, 0.707$\pm$0.150 in V and 
0.575$\pm$0.178 in I.

     The results presented above are essentially identical to our preliminary solutions
obtained from a subsample of 11 SNe with a similar color cutoff (equations 11 and 12 of
Paper IV), namely,  0.847$\pm$0.128 in B and 0.787$\pm$ 0.109 in V.  The inclusion of the 
remaining Cal\'{a}n/Tololo SNe has significantly populated the
absolute magnitude/decline rate diagram in the range 0.8 $<$
$\Delta$m$_{15}$(B) $<$ 1.7,  strongly confirming 
the reality of the magnitude-decline rate relationship.

In Figure 2 the nearby sample of SNe Ia is shown superimposed on the Cal\'{a}n/Tololo
points (excluding the two possibly reddened SNe 90Y and 93H). The combined data
set is generally consistent with the relations shown
in Figure 1 and Table 3.   As discussed in Paper IV,  the Cal\'{a}n/Tololo slopes are 
significantly smaller than Phillips' original values, but much of this change
comes about by restricting the sample in color. As
also noted by Phillips, the slope of the absolute magnitude/decline rate 
relationship becomes smaller toward longer wavelengths. The differences seen
in Figure 2 between the samples consist of two objects (SNe 92A and 91bg)
being significantly fainter than the mean of the distant sample at the same $\Delta$m$_{15}$(B).
Based on the color criterion adopted above, SN 91bg would be
excluded from the fits, but SN 92A would remain.
While the photometry of the nearby sample is probably better than our distant sample, it is
possible that the relative distances obtained from the Hubble law are more reliable
than the SBF/PNLF distances used for the individual objects in the nearby sample,
since the peculiar motions of the distant galaxies are a small fraction of their redshifts.
We note that the nearby SNe with $\Delta$m$_{15}$(B) $> 1.2$ have distance moduli derived
exclusively from SBF/PNLF, whereas all of the events with $\Delta$m$_{15}$(B) $< 1.2$
have Cepheid-based distances.
Hence, any systematic differences between the two
techniques would lead to a different slope compared to that of the distant sample.
On the other hand, the Cal\'{a}n/Tololo sample suffers from a range of biases, including
Malmquist effects; SNe fainter than M$^B_{\rm{MAX}}$ $\simeq$ --18.1
would not be seen in more than 1/2 the volume
of the survey (and indeed the faint  SN 92K is the lowest redshift object in
the sample). Thus the true SNe Ia luminosity function remains to be determined
and the small differences seen in Figure 2 have several possible explanations,
although their effect on the derivation of the Hubble constant is probably small
(see Paper VI).

\section{  Morphological Dependencies}

     In paper IV we investigated the relation between the luminosities of the SNe and their
environments, and noticed that the most luminous objects were hosted by spiral galaxies. In
that paper we separated the galaxies in two basic categories, namely, spirals and nonspirals.
Given the larger number of SNe included in this paper we have considered a finer grid of
galaxy types: E, S0, Sa, Sb, Sc, and Irr. As can be seen in Table 1, in some cases the
classification was intermediate between two categories, and in a few cases we were not able to
provide an adequate classification (due to the distances of the galaxies). 
Figure 3 shows the absolute BVI magnitudes of the SNe listed in Table 1
and Table 2, plotted as a function of the morphological types of their
parent galaxies.  We recover our previous result in the sense that the
brightest SNe occur in late-type galaxies, although admittedly the effect
is stronger in the nearby sample than in the Cal\'{a}n/Tololo SNe Ia.  Note 
that this feature is not the result of uncorrected dust absorption, which 
would be expected to make SNe in spirals less luminous on average.

We find an even more striking relationship between the decline rate of
the B light curves and the morphological classifications of the host galaxies.
As shown in Figure 4, SNe in spirals span a wide range in decline rates,
whereas elliptical galaxies have not produced slow-decline SNe.  Note that reddening has
no effect on the points in this diagram.  These results
suggest that galaxies with a younger stellar population host the intrinsically
brightest SNe Ia (slowest decliners).
In this sense, the Cal\'{a}n/Tololo database provides a challenging opportunity to confront
observations with a theory that can predict this morphological dependence
of the SNe parameters.

\section{Discussion and Conclusions}

The application of the peak magnitude-decline rate relation permits a  significantly
reduced scatter in the SNe Ia
Hubble diagram, to levels of 0.13$^m$--0.17$^m$. Without use of this relation,
the mean magnitudes and dispersions for the Cal\'{a}n/Tololo  sample (from Table 1)  are:
--19.05$\pm$0.38 in B, --19.12$\pm$0.26 in V and --18.91$\pm$0.20 in the I band.
The reduction of $\sigma$ by approximately a factor of 2  permits  SNe Ia to be
used as excellent distance indicators 
(with precisions in relative distances $\sim$7-10\%).

Recent work by Nugent \etal (1995) suggests that the correlation between 
peak luminosity and decline rate (Fig. 1) is related to a systematic variation in
effective temperature, presumably corresponding to the amount of $^{56}$Ni produced
in the explosion.  The origin of the relations
between luminosity (Fig. 3)  or decline rate (Fig. 4)  and host galaxy morphological type are
not understood at this point and need further theoretical exploration
and modelling. The linear approximation to the luminosity-decline rate
relation is all that is justified by the current data. Further study of
the abnormal color SNe (e.g.,  SN 91bg and SN 92K) may provide clues to aspects 
of these relations.

A recent paper by Tammann and Sandage (1995; hereinafter T\&S) calls into 
question the reality of the decline rate relation and any morphological
dependences on SNe luminosity.  The T\&S version includes 10 SNe with ``modern data'' 
    selected from Phillips (1993) and Paper IV, four 
    SNe in Virgo cluster galaxies, and three other random SNe with v $>$ 1100 
    \kms for which T\&S provide their own estimates of the decline rate 
    parameter $\Delta$m$_{15}$(B).  The 10 SNe with modern data show clear evidence for 
    a correlation, which is not surprising since 7 of these are drawn from 
    Paper IV.  The four Virgo events show much more scatter, 
    but this is also to be expected since no 
    allowance has been made for the considerable depth of the cluster.  The 
    three additional SNe (1970J, 1975N, 1976J) included in the diagram by 
    T\&S introduce further scatter but, again, this is not surprising since 
    the data for these events is greatly inferior to that of the SNe with 
    ``modern data''.  Indeed, by including the latter 3 SNe (as well as 1984A 
    from the Virgo subsample), T\&S contradict their own statement 
    that ``the photometric data must be exquisite'' to make a precise 
    determination of $\Delta$m$_{15}$(B).

 From Figure 2 of their paper, T\&S conclude that the steepest allowable 
    slopes for the decline rate-luminosity relation are 0.88 in B and 0.75 
    in V.  Although their statement that these values are ``factors of 3.1 
    and 2.6 smaller than the original Phillips (1993) result'' is technically
    correct, it is important to realize that these steeper slopes were
    due largely to the inclusion of the low-luminosity event SN 1991bg
    which T\&S specifically exclude from their sample on the grounds of 
    spectroscopic peculiarity.  Even more confusing, however, is T\&S's 
    statement that ``the adopted slope values by Hamuy \etal of 1.62 in B 
    and 1.76 in V are also larger than ours by an average factor of two.''
    In fact, Hamuy \etal give three different estimates for the slopes in 
    B and V.  The values quoted by T\&S were derived from the 6 events in the 
    Phillips (1993) nearby sample with $\Delta$m$_{15}$(B)$ <$ 1.5 and have large associated
    errors (1.624 $\pm$ 0.582 and 1.764$\pm$0.570); the other estimates are
    1.365$\pm$ 0.275 in B and 1.142 $\pm$ 0.240 in V obtained from the Cal\'{a}n/Tololo
    SNe Ia with $\Delta$m$_{15}$(B) $<$ 1.5, and 0.847$\pm$0.128 in B and 0.787$\pm$ 0.109
    in V derived from the Cal\'{a}n/Tololo sample excluding only the 1991bg-like
    SN 1992K.  The slopes derived by T\&S should, of course, be compared 
    with the latter values -- not surprisingly (since 7 of the 10 SNe
    in the T\&S ``modern data'' sample are taken from Paper IV), they are
    very similar.

The new data presented in this paper both supports the reality of the
decline rate-luminosity relation and strengthens the evidence for a dependence
of the luminosity on the morphological type of the host galaxy. The
conclusions of the T\&S paper are based on confused comparisons with
the results of our earlier papers and the addition of old, imprecise 
photographic data which dilutes and obscures the relations seen with higher-quality modern data.

                              Acknowledgments

We are grateful to George Jacoby for several interesting discussions.
This paper has been possible thanks to grant 92/0312 from Fondo Nacional de Ciencias y
Tecnolog\'{i}a (FONDECYT-Chile).
MH acknowledges support provided for this work by the National Science Foundation
through grant number GF-1002-95 from the Association of Universities for Research
in Astronomy, Inc., under NSF Cooperative Agreement No. AST-8947990.
JM and MH acknowledge support provided from C\'{a}tedra Presidencial 1996-1997.
Thanks to A. Filippenko and T. Matheson for allowing us to
examine their spectrum of SN 1990Y.

\eject

\centerline{                                References}

\noindent Barbon, R., Ciatti, F., \& Rosino, L. 1973, A\&A, 25, 241

\noindent Barbon, R., Ciatti, F., \& Rosino, L. 1982, A\&A, 116, 35

\noindent Boisseau, J.R., \& Wheeler, J.C. 1991, AJ, 101, 1281

\noindent Burstein, D., \& Heiles, C. 1982, AJ, 87, 1165

\noindent Burstein, D., \& Heiles, C. 1984, ApJS, 54, 33

\noindent Buta, R.J., \& Turner, A. 1983, PASP, 95, 72

\noindent Ciardullo, R.B., Jacoby, G.H., \& Tonry, J.L. 1993, ApJ, 419, 479

\noindent Filippenko, A., \& Shields, J.C. 1990, IAU Circ., No. 5083

\noindent Filippenko, A.V., \etal 1992, AJ, 104, 1543

\noindent Hamuy, M., Phillips, M.M, Maza, J., Wischnjewsky, M., Uomoto, A.,
      Landolt, A.U., \& Khatwani, R. 1991, AJ, 102, 208

\noindent Hamuy, M., \etal 1993a, AJ, 106, 2392 (Paper I)

\noindent Hamuy, M., Phillips, M.M., Wells, L.A., \& Maza, J. 1993b, PASP, 105, 787

\noindent Hamuy, M., \etal 1994, AJ, 108, 2226 (Paper III)

\noindent Hamuy, M., Phillips, M.M., Maza, J., Suntzeff, N.B., Schommer, R.A.,
     \& Avil\'{e}s, R. 1995, AJ, 109, 1 (Paper IV)

\noindent Hamuy, M., Phillips, M.M., Suntzeff, N.B., Schommer, R.A., Maza, J,
     \& Avil\'{e}s, R. 1996a, AJ, this volume (Paper VI)

\noindent Hamuy, M., \etal 1996b, AJ, this volume (Paper VII)

\noindent Hamuy, M., Phillips, M.M., Suntzeff, N.B., Schommer, R.A., Maza, J,
     Smith, R.C., Lira, P., \& Avil\'{e}s, R. 1996c, AJ, this volume (Paper VIII)

\noindent Hough, J.H., Bailey, J.A., Rouse, M.F., \& Whittet, C.B. 1987, MNRAS, 227, 1P.

\noindent Jacoby, G.H. 1996, in proceedings of {\it The Extragalactic Distance Scale}, 
     STScI Spring meeting, ed. M. Livio, in preparation

\noindent Leibundgut, B., \etal 1993, AJ, 105, 301

\noindent Lira, P. 1995, MS THESIS, Universidad de Chile.

\noindent Maza, J., Hamuy, M., Phillips, M.M., Suntzeff, N.B., \& Avil\'{e}s, R. 1994,
     ApJ, 424, L107 (Paper II)

\noindent Nugent, P., Phillips, M., Baron, E., Branch, D., \& Hauschildt, P. 1995, ApJ, 455, L147

\noindent Phillips, M.M., \etal 1987, PASP, 99, 592

\noindent Phillips, M.M. 1993, ApJ, 413, L105

\noindent Phillips, M.M., \& Eggen, O.J. 1996, in preparation

\noindent Pierce, M.J., \& Jacoby, G.H. 1995, AJ, 110, 2885

\noindent Press, W.H., Teukolsky, S.A., Vetterling, W.T., \& Flannery, B.P. 1992,
     {\it Numerical Recipes in Fortran, Second Edition} (Cambridge University 
     Press, Cambridge), pp. 660-664

\noindent Pskovskii, Y.P. 1977, SvA, 21, 675

\noindent Pskovskii, Y.P. 1984, SvA, 28, 658

\noindent Richmond, M.W., \etal 1995, AJ, 109, 2121

\noindent Saha, A., Labhardt, L., Schwengeler, H., Macchetto, F.D., Panagia, N.,
     Sandage, A., \& Tammann, G.A. 1994, ApJ, 425, 14

\noindent Saha, A., Sandage, A., Labhardt, L., Schwengeler, H., Tammann, G.A., 
     Panagia, N., \& Macchetto, F.D. 1995, ApJ, 438, 8

\noindent Saha, A., Sandage, A., Labhardt, L., Tammann, G.A., Macchetto, F.D.,
     \& Panagia, N. 1996, ApJ, in press

\noindent Sandage, A. 1961, {\it The Hubble Atlas of Galaxies} (Carnegie Institution of 
     Washington, Washington DC)

\noindent Sandage, A., \& Tammann, G.A. 1981, {\it A Revised Shapley-Ames Catalog of Bright
     Galaxies} (Carnegie Institution of Washington, Washington DC)

\noindent Sandage, A., Saha, A., Tammann, G.A., Labhardt, L., Panagia, N., \&
     Macchetto, F.D. 1996, ApJ, 460, L15

\noindent Smith, R.C., \etal 1996, in preparation

\noindent  Suntzeff, N.B. 1996, in {\it Supernovae and Supernova Remnants}, IAU Colloquium 145,
     ed. R. McCray (Cambridge University Press, Cambridge), in press

\noindent Tammann, G.A., \& Sandage, A. 1995, ApJ, 452, 16 (T\&S)

\noindent Tonry, J.L. 1996, in proceedings of {\it The Extragalactic Distance Scale}, 
     STScI Spring meeting, ed. M. Livio, in preparation

\noindent Tsvetkov, D.Y. 1982, SvAL, 8, 115

\noindent Vaughan, T.E., Branch, D., Miller, D.L., \& Perlmutter, S. 1995,
ApJ, 439, 558. 

\eject

\centerline{                             Figure Captions}

Figure 1. The absolute B, V and I magnitudes (from Table 1) of the 29 Cal\'{a}n/Tololo SNe Ia plotted as a
function of $\Delta$m$_{15}$(B). Points with dotted error bars correspond
to the two SNe (90Y and 93H) suspected to be significantly reddened by dust and to the
intrinsically red SN 1992K. The ridge lines
correspond to weighted linear least-squares fits to the remaining 26 (B \& V) and 22 (I) SNe with
0.87 $\leq$ $\Delta$m$_{15}$(B) $\leq$ 1.69.

Figure 2. As in Fig 1, but with the the open circles corresponding to the
Cal\'{a}n/Tololo sample (excluding the two possibly reddened SNe 90Y and 93H),
and the filled circles to the nearby sample (from Table 2). 

Figure 3. The absolute B, V and I  magnitudes of the Cal\'{a}n/Tololo
SNe Ia (open circles) and the nearby SNe (filled circles) plotted as
a function of the morphological types of their host galaxies.

Figure 4. The decline rate of the B light curve [$\Delta$m$_{15}$(B)] of
both the Cal\'{a}n/Tololo (open circles) and nearby (filled circles)
samples of  SNe Ia plotted
as a function of the morphological types of their host galaxies.

\evensidemargin=0.0in
\oddsidemargin= 0.0in
\small

\begin{tabular}{lcccccll}
\multicolumn{8}{c}{\bf Table 1. Colors and Absolute Magnitudes of the Cal\'{a}n/Tololo Supernovae Ia} \\
&&&&&&&\\
(1) & (2) & (3) & (4) & (5) & (6) & (7) & (8)\\
&&&&&&&\\
\hline\hline\\
                                                                                   
 SN & B$_{\rm{MAX}}-$V$_{\rm{MAX}}$ & M$^B_{\rm{MAX}}$ &  M$^V_{\rm{MAX}}$ &  M$^I_{\rm{MAX}}$ &
$\Delta$m$_{15}$(B)& (B-V)$_{gal}$ & Galaxy \\
      & &           +5log(H$_0$/65)& +5log(H$_0$/65)& +5log(H$_0$/65) & & $\pm$0.10&     Type\\
& & & & & & & \\
\hline\\

90O  & 0.01(05) & -19.40(17)  & -19.41(16)  & -19.02(17)  & 0.96(10)  & 0.98  & SBa  \\ 
90T  & 0.04(10) & -19.17(24)  & -19.21(19) &  -18.98(19)  & 1.15(10)  & 0.71  & Sa   \\
90Y  & 0.33(10) & -18.56(24)  & -18.89(20)  & -18.65(19)  & 1.13(10)  & 0.56  & E?   \\
90af &  0.05(03) &  -18.95(11) & -19.00(11) &     ---      & 1.56(05)  & 0.92  & SB0   \\
91S  & 0.04(10)  & -19.24(22)  & -19.28(18)  & -18.98(17)  & 1.04(10)  & 0.69  & Sb   \\
91U  & 0.06(10)  & -19.49(25)  & -19.55(21)  & -19.37(20)  & 1.06(10)  & 0.71  & Sbc   \\
91ag &  0.08(05) &  -19.40(35) & -19.48(35) &  -19.16(37) &  0.87(10) &  0.46 &  SBb   \\
92J  & 0.12(10)  & -18.92(23)  & -19.04(19)  & -18.78(18)  & 1.56(10)  & 0.95  & E/S0   \\
92K  & 0.74(10)  & -17.72(44)  & -18.46(42)  & -18.61(42)  & 1.93(10)  & 0.86  & SBb   \\
92P  & -0.03(03) &   -19.34(18) &-19.31(18) &  -19.03(18) &   0.87(10) &  0.73 &  SBa   \\
92ae &  0.11(05) &   -19.07(13) &-19.18(10) &    ---     &  1.28(10) &  0.53 &  E1?   \\
92ag &  0.13(05) &   -18.98(19) &-19.11(18) &  -18.98(18) &  1.19(10) &  0.68 &  S   \\
92al &  -0.05(03) &  -19.47(32) &-19.42(31) &  -19.13(31) &  1.11(05) &  0.60 &  Sb   \\
92aq &  0.10(05)  & -18.89(10)  &-18.99(08)  & -18.57(10)  & 1.46(10)  & 1.02 &  Sa?   \\
92au &  0.05(10)  & -19.03(22)  &-19.08(18)  & -18.83(17)  & 1.49(10)  & 0.87 &  E1   \\
92bc &  -0.08(03) &  -19.64(23) &-19.56(23) &  -19.22(22) &  0.87(05) &  0.66 &  Sab   \\
92bg &  -0.04(05) &  -19.36(15) &-19.32(14) &  -19.04(14) &  1.15(10) &  0.65 &  Sa   \\
92bh &  0.08(05)  & -18.89(13)  &-18.97(11)  & -18.79(11)  & 1.05(10)  & 0.50  & Sbc   \\
92bk &  0.00(05)  & -19.03(12)  &-19.03(10)  & -18.83(10)  & 1.57(10)  & 0.96  & E1   \\
92bl &  0.00(05)  & -19.13(13)  &-19.13(12)  & -18.85(12)  & 1.51(10)  & 0.86  & SB0/SBa   \\
92bo &  0.01(03)  & -18.76(25)  &-18.77(25)  & -18.65(24)  & 1.69(05)  & 0.97  & E5/S0   \\
92bp &  -0.05(05) &  -19.40(09) &-19.35(08) &  -19.03(08) &  1.32(10) &  0.72 &  E2/S0   \\
92br &  0.04(05)  & -18.66(18)  &-18.70(11)  &   ---       & 1.69(10)  & 0.97  & E0   \\
92bs &  0.07(05)  & -18.96(11)  &-19.03(10)  &   ---       & 1.13(10)  & 0.51  & SBb   \\
93B  & 0.12(05)   & -19.04(13)  &-19.16(11)  & -18.87(12)  & 1.04(10)  & 0.53  & SBb   \\
93H  & 0.23(05)   & -18.45(19)  &-18.68(19)  & -18.74(19)  & 1.69(10)  & 0.84  & SBb(rs)   \\
93O  & -0.09(03)  & -19.23(11)  &-19.14(10)  & -18.91(10)  & 1.22(05)  & 0.66  & E5/S01   \\
93ag &  0.03(05)  & -19.10(12)  &-19.13(11)  & -18.81(11)  & 1.32(10)  & 0.80  & E3/S01   \\
93ah &  -0.04(10) &  -19.28(26) &-19.24(22) &  -18.93(21) & 1.30(10)  & 0.83  & S02   \\
\hline\hline\\
             
\end{tabular}
\pagestyle{empty}

\tiny
\evensidemargin=-0.5in
\oddsidemargin=-0.5in
\begin{tabular}{ccccccccccll}
\multicolumn{12}{c}{\bf Table 2. Nearby SNe Sample     }\\
&&&&&&&&&&&\\
(1) & (2) & (3) & (4) & (5) & (6) & (7) & (8) & (9) & (10) & (11) & (12)\\
&&&&&&&&&&&\\
\hline\hline\\
&&&&Distance&&&&&&\\
 SN & Host & Morph. & Distance  & Modulus & Method & $\Delta$m$_{15}$  & E(B-V)& M$^B_{\rm{MAX}}$& M$^V_{\rm{MAX}}$ &  M$^I_{\rm{MAX}}$ & Photometry   \\
&Galaxy& Type &Modulus& References$^{a}$&&&&&&&References$^{b}$\\
&&&&&&&&&&&\\
\hline
&&&&&&&&&&&\\
37C & IC 4182  & S/Irr & 28.36(09) & 1 & Cepheids & 0.87(10) & 0.00(02) & -19.56(15) & -19.54(16) &  & 1 \\
72E & NGC 5253 & Irr  & 27.97(07) & 2 & Cepheids & 0.87(10) & 0.05(02) & -19.69(18) & -19.64(18) & -19.26(21) & 2 \\
80N & NGC 1316 & Sa$_{pec}$ & 31.23(06) & 3,4 & SBF/PNLF & 1.28(04) & 0.00(02) & -18.74(11) & -18.79(09) & -18.53(08) & 3 \\ 
81B & NGC 4536 & Sc & 31.10(13) & 5 & Cepheids & 1.10(05) & 0.00(02) &  -19.07(16) & -19.17(15) &  & 4,5,6 \\
86G & NGC 5128 & SO+S$_{pec}$ &  27.86(05) & 3,4 & SBF/PNLF & 1.73(07) & 0.65(10) &   -18.08(42) & -18.43(32) & -18.45(22) & 7 \\ 
90N & NGC 4639 & Sb &  32.00(23) & 6  & Cepheids & 1.07(05)  & 0.00(02)  &  -19.26(25) & -19.28(24) & -19.05(24) & 8 \\ 
91bg& NGC 4374 & E1 & 31.26(05) & 3 & SBF & 1.93(10) &  0.03(02) &   -16.62(14) & -17.38(09) & -17.81(08) & 9,10 \\  
92A & NGC 1380 & SO/Sa &  31.00(10) & 3 & SBF &  1.47(05) &  0.00(02)&   -18.43(13) & -18.45(12) & -18.20(11) & 11 \\ 
94D & NGC 4526 & SO & 30.86(08) & 3 & SBF &  1.32(05) &  0.00(02)&   -19.00(12) & -18.96(11) & -18.75(09) & 12 \\ 
&&&&&&&&&&&\\
\hline\hline\\
\end{tabular}

 $^{a}$References.-
 $^{1}$Saha \etal (1994);
 $^{2}$Saha \etal (1995);
 $^{3}$Tonry (1996);
 $^{4}$Ciardullo \etal (1993);
 $^{5}$Saha \etal (1996);
 $^{6}$Sandage \etal (1996).

 $^{b}$References.-
 $^{1}$Pierce \& Jacoby (1995);
 $^{2}$Phillips \& Eggen (1996);
 $^{3}$Hamuy \etal (1991);
 $^{4}$Buta \& Turner (1983);
 $^{5}$Barbon \etal (1982);
 $^{6}$Tsvetkov (1982);
 $^{7}$Phillips \etal (1987);
 $^{8}$Lira (1995);
 $^{9}$Filippenko \etal (1992);
 $^{10}$Leibundgut \etal (1993);
 $^{11}$Suntzeff \etal (1996);
 $^{12}$Smith \etal (1996).

\small
\eject          
\evensidemargin=-0.5in
\oddsidemargin=-0.5in
\begin{tabular}{ccccccl}
\multicolumn{7}{c}{\bf Table 3.  Least-Squares Fits     }\\
&&&&&&\\
\multicolumn{7}{c}{   M$_{MAX}$ = a + b[$\Delta$m$_{15}$(B) - 1.1] } \\
&&&&&&\\
\hline\hline\\
  Bandpass &          a    &         b   &    $\sigma(mag)$  & \chinusq& n &    sample\\
&&&&&&\\
\hline                                                                            
&&&&&&\\
         
         B      &   -19.258(0.048) & 0.784(0.182) & 0.17 & 1.24  & 26 & ``low extinction''\\
                &   -19.256(0.053) & 0.860(0.210) & 0.20 & 1.71  & 18 & peak subsample\\
&&&&& &    \\
         V      &   -19.267(0.042) & 0.707(0.150) & 0.14 & 0.97  & 26 & ``low extinction''\\
                &   -19.254(0.046) & 0.743(0.173) & 0.15 & 1.27  & 18 &  peak subsample\\
&&&&&  &        \\
         I      &   -18.993(0.044) & 0.575(0.178) & 0.13 & 0.85  & 22 &``low extinction''\\
                &   -18.975(0.049)&  0.654(0.231) & 0.13 & 1.02  & 14 & peak subsample \\
\hline\hline\\
          
\end{tabular}

\eject
\begin{figure}
\psfull
\centerline{\hbox{
\psfig{figure=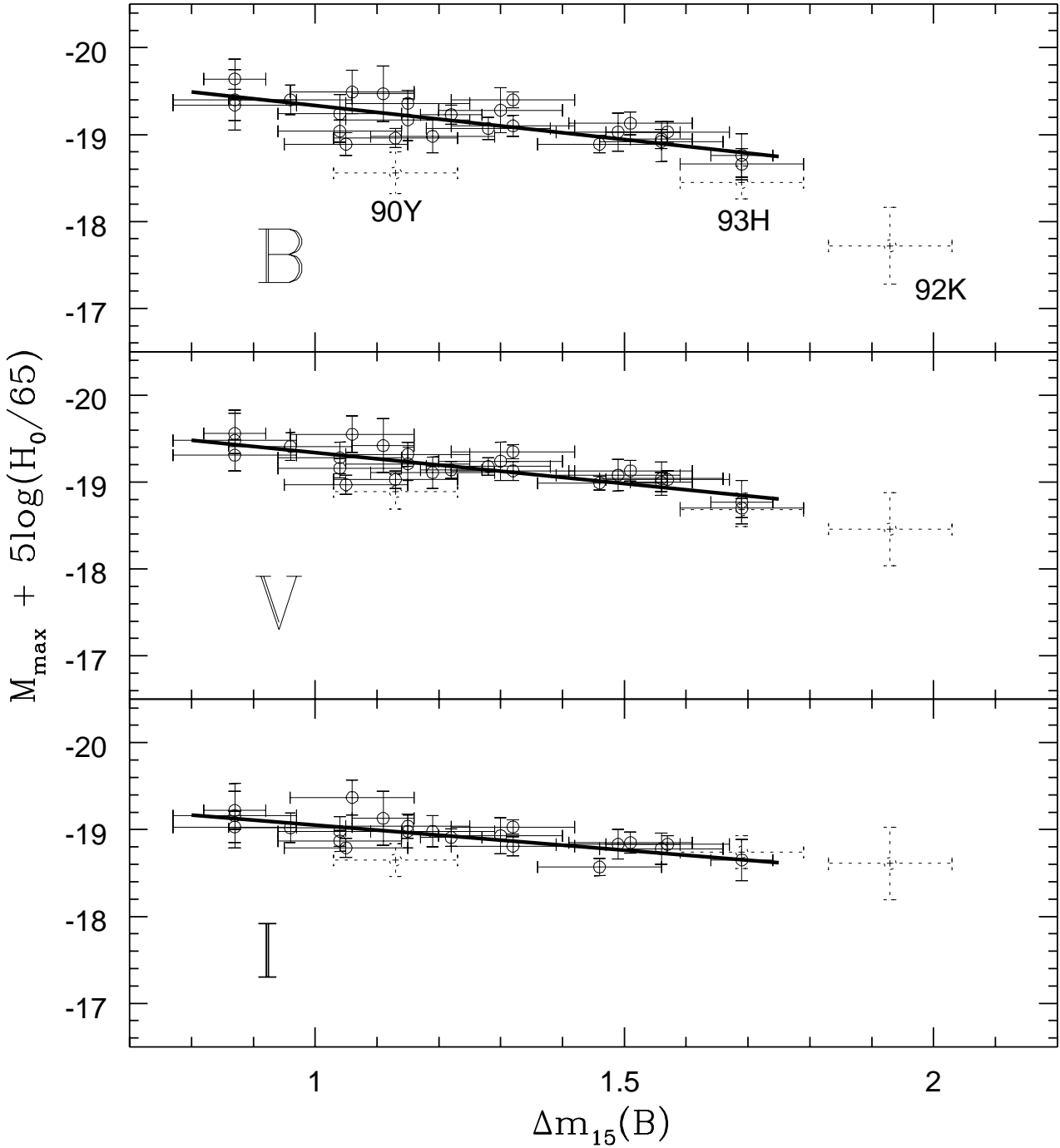}
}}
\caption{The absolute B, V and I magnitudes (from Table 1) of the 29 Cal\'{a}n/Tololo SNe Ia plotted as a
function of $\Delta$m$_{15}$(B). Points with dotted error bars correspond
to the two SNe (90Y and 93H) suspected to be significantly reddened by dust and to the
intrinsically red SN 1992K. The ridge lines
correspond to weighted linear least-squares fits to the remaining 26 (B \& V) and 22 (I) SNe with
0.87 $\leq$ $\Delta$m$_{15}$(B) $\leq$ 1.69.}
\end{figure}

\eject
\begin{figure}
\psfull
\centerline{\hbox{
\psfig{figure=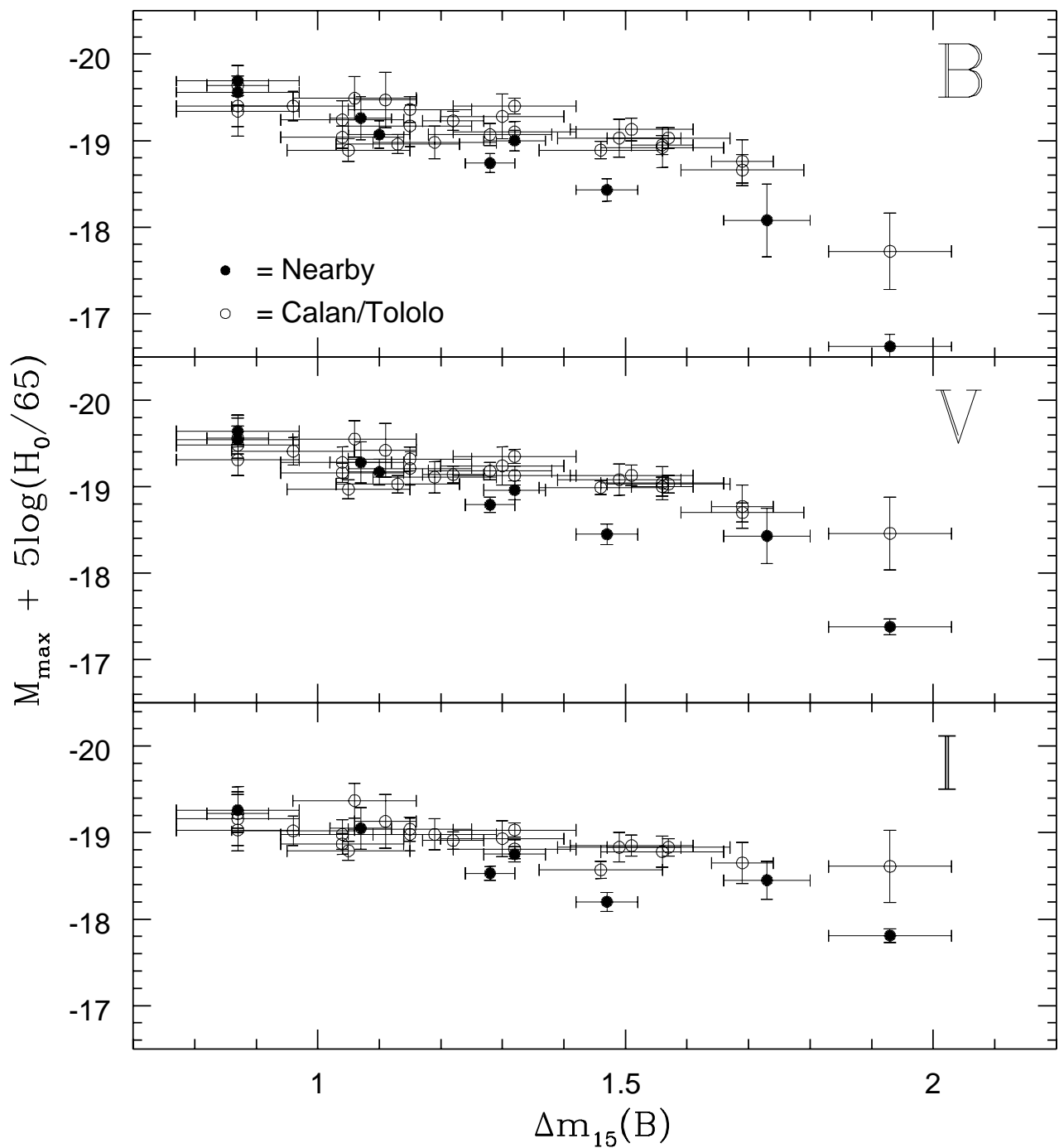}
}}
\caption{As in Fig 1, but with the the open circles corresponding to the
Cal\'{a}n/Tololo sample (excluding the two possibly reddened SNe 90Y and 93H),
and the filled circles to the nearby sample (from Table 2).}
\end{figure}

\eject
\begin{figure}
\psfull
\centerline{\hbox{
\psfig{figure=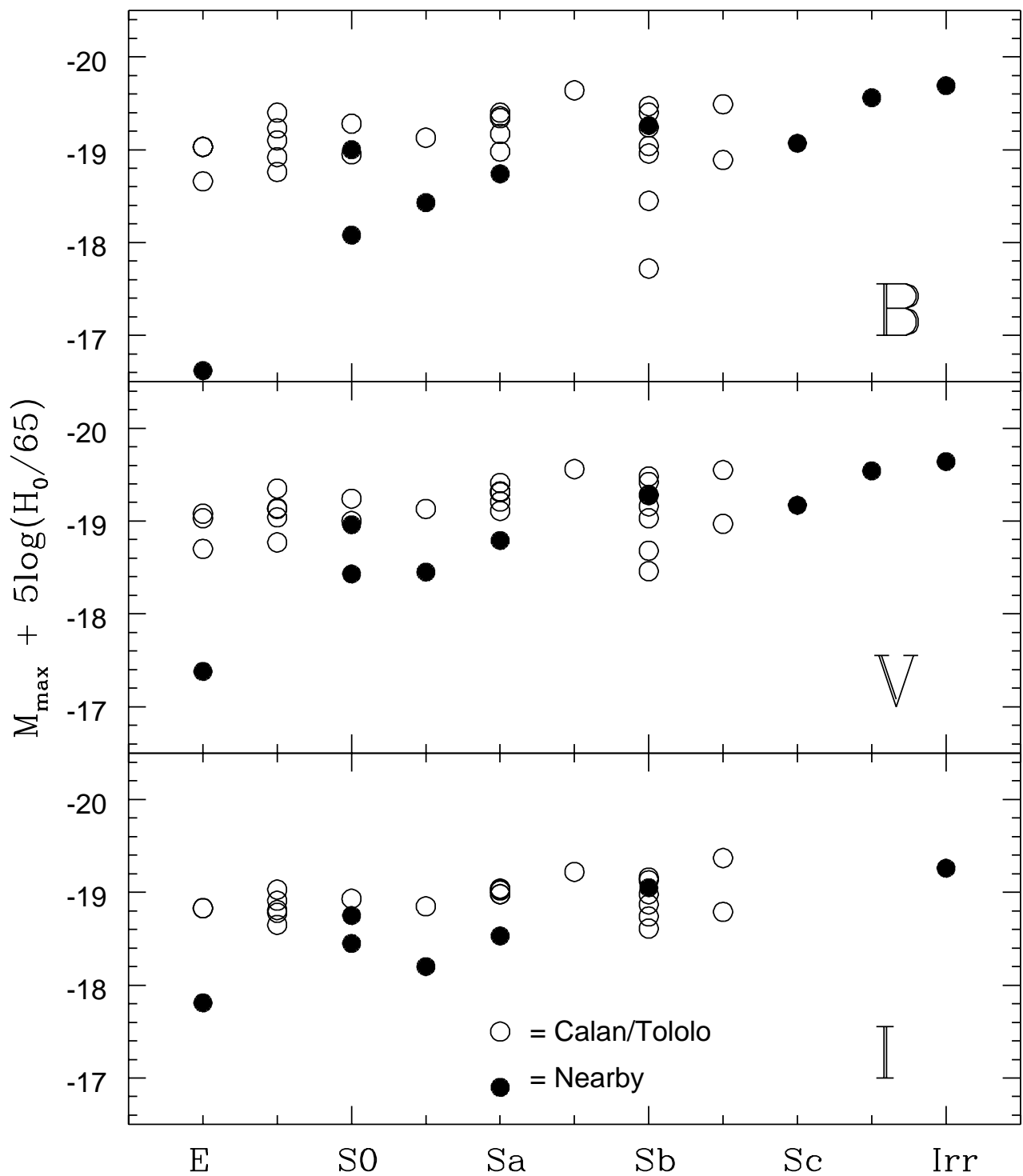}
}}
\caption{The absolute B, V and I  magnitudes of the Cal\'{a}n/Tololo
SNe Ia (open circles) and the nearby SNe (filled circles) plotted as
a function of the morphological types of their host galaxies.}
\end{figure}

\eject
\begin{figure}
\psfull
\centerline{\hbox{
\psfig{figure=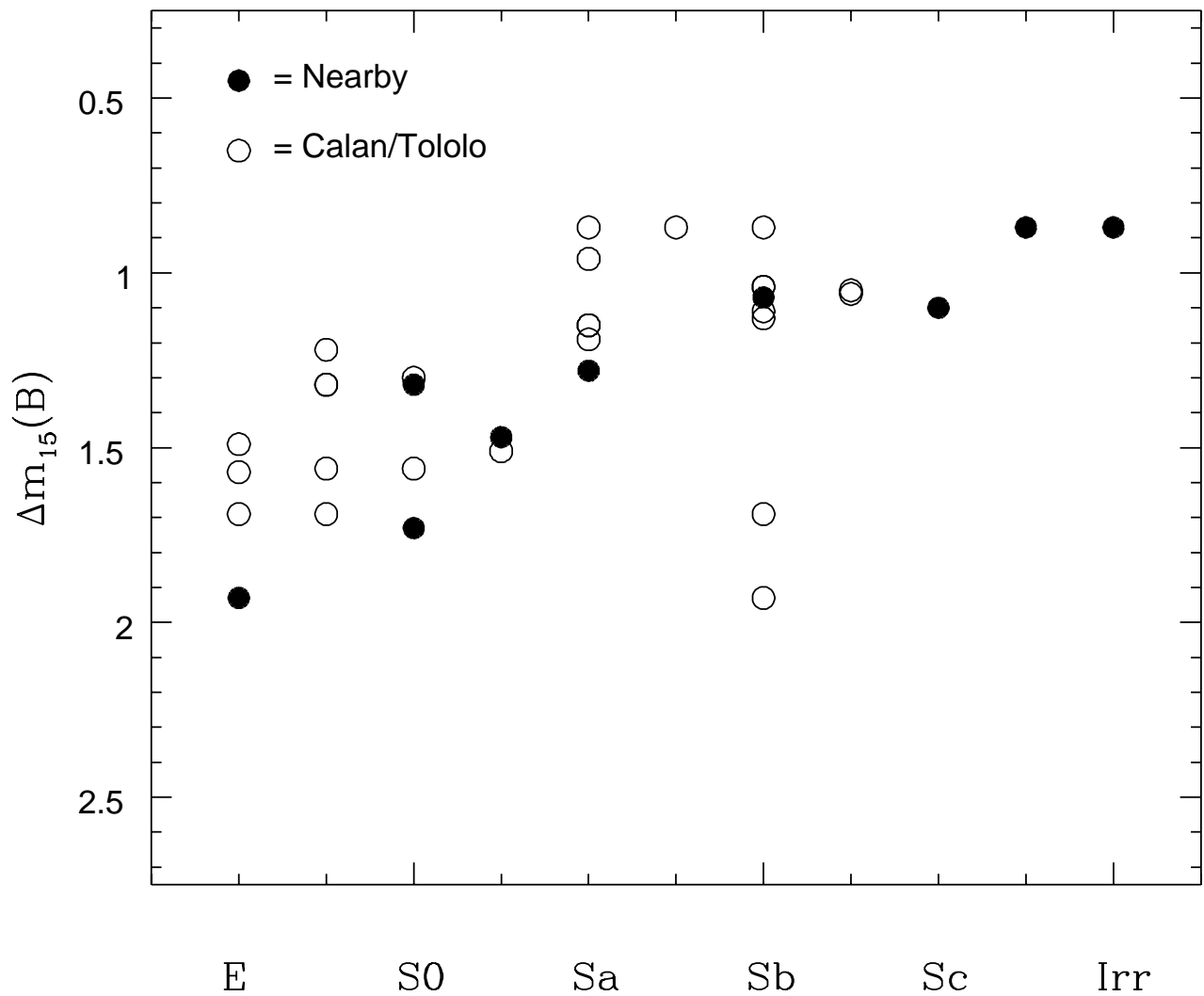}
}}
\caption{The decline rate of the B light curve [$\Delta$m$_{15}$(B)] of
both the Cal\'{a}n/Tololo (open circles) and nearby (filled circles)
samples of  SNe Ia plotted
as a function of the morphological types of their host galaxies.}
\end{figure}

\end{document}